\documentclass[aps,prd,showpacs,preprint,nofootinbib]{revtex4-1}
\usepackage{amsmath,amsthm,amssymb,verbatim,graphicx,color,fullpage}
\allowdisplaybreaks

\usepackage[colorlinks]{hyperref}

\long\def\symbolfootnote[#1]#2{\begingroup%
\def\thefootnote{\fnsymbol{footnote}}%
\renewcommand{\baselinestretch}{0.9}\footnote[#1]{#2}\endgroup}

\newcommand{\SO}{\mathrm{SO}}
\newcommand{\SU}{\mathrm{SU}}
\newcommand{\U}{\mathrm{U}}
\newcommand{\QCD}{\mathrm{QCD}}
\newcommand{\fo}{\mathrm{fo}}
\newcommand{\hi}{\mathrm{hi}}
\newcommand{\lo}{\mathrm{lo}}
\newcommand{\Pl}{\mathrm{Pl}}
\newcommand{\EW}{\mathrm{EW}}
\newcommand{\ann}{\mathrm{ann}}
\newcommand{\GeV}{\mathrm{GeV}}
\newcommand{\THDM}{\mathrm{2HDM}}

\newcommand{\Tr}{\mathrm{Tr}\,}

\newcommand{\tPhi}{\tilde{\Phi}}

\newcommand{\vev}[1]{\langle #1 \rangle}

\begin{document}
{\large {\bf
\begin{center}
Gravitational wave signature of generic disappearance of \\
$Z_2$-symmetry breaking domain walls
\end{center}
}}

\vskip 1.0 cm

\centerline{
\bf Piyali Banerjee\symbolfootnote[1]{
\renewcommand{\baselinestretch}{0.8}
School of Engineering and Technology, Nagaland University,
Dimapur 797112, India. 
Part of this work was done while the author was
a DST WOS-A PI at IIT Bombay.
{\sf banerjee.piyali3@gmail.com}
}
and
Urjit A. Yajnik\symbolfootnote[2]{
Department of Physics, Indian Institute of 
Technology Bombay, Mumbai 400076, India.
{\sf yajnik@iitb.ac.in}
}
}

\bigskip
\bigskip

\begin{center}
{\large \bf Abstract}
\end{center}

\bigskip

Breaking of discrete parity at high scale gives  rise to  $Z_2$-domain 
walls (DW). 
The metastability of such walls can make them relatively long lived and 
contradict standard cosmology. 
We consider two classes of theories with similar underlying feature, 
the left right 
symmetric  theories and two Higgs doublet models. Both of them
possess some breaking of $Z_2$ discrete symmetries. As a first step,
domains form at a high energy scale during parity breaking. In the second
step, these domains
further decompose into subdomains due to $Z_2$ symmetry breaking in
two Higgs doublet models closer to the electroweak scale. 
We show that after this two step formation of domains and subdomains,
a QCD instanton induced energy difference
can remove the domain walls as well as the subdomain walls at
around the same time successfully. 
The removal occurs purely as the result of a chance event
taking place with probability very close to 0.25, and does not require one
to introduce any non-renormalisable $Z_2$-symmetry breaking term to
the Lagrangian.
We then investigate the gravitational
waves arising from the collapse of such domain walls and 
show that 
the peak frequency of these waves lies in the 
$10^{-7}$--$10^{-6}~\mbox{Hz}$ band, corresponding to annihilation 
temperatures of $1$--$10$ GeV. This frequency band is sensitive  
to pulsar timing array based experiments such as SKA and NANOGrav. The
recent NANOGrav results rule out our DW collapse model for higher
values of parity breaking scale above $10^7$ GeV. Our DW collapse model
with parity breaking scales below $10^7$ GeV  remains consistent with 
the current NANOGrav results and has a good
chance of being seriously tested in future pulsar timing based experiments.

\bigskip

PACS Numbers: 
\vfill 

\section{Introduction}
Several Beyond Standard Model (BSM) theories of particle physics 
have been proposed over the years to address various issues that the
Standard Model (SM) cannot explain. Examples of these issues include
the existence of small non-zero neutrino masses, necessity of more
sources of CP violation and more sources of first order phase 
transition as required for successful baryogenesis, 
hints of unification of couplings at high energy etc.
Many of these models 
possess a common feature that we will refer to as
$Z_2$-symmetry breaking in this paper. In $Z_2$-symmetry breaking,
the models possess an accidental or deliberate $Z_2$-symmetry at
high energy which gets broken at lower energies due to spontaneous
symmetry breaking. Breaking of $Z_2$-symmetry gives rise to energy
barriers called domain walls (DW) which contradict the cosmology that
we observe today. In this paper we study this issue and propose a
mechanism to remove them successfully in the early universe itself for
many models, 
without invoking any new non-renormalisable physics. We then discuss 
prospects of direct verification of our proposed mechanism in 
upcoming experiments.

A famous class of models possessing $Z_2$-symmetry breaking are the
left right symmetric extensions of the Standard Model (LRSM). 
First proposed by Mohapatra and Senjanovi\'{c}
\cite{MohapatraLR}, LRSM is a minimal extension of SM
based on the gauge group 
$\SU(3)_c \times \SU(2)_L \times \SU(2)_R \times \U(1)_{B-L}$ augmented
with a discrete parity or left-right symmetry $P$ which
interchanges the gauge bosons and
triplet Higgs fields of $\SU(2)_L$ and $\SU(2)_R$ besides implementing
space-time parity. 
The usual $\SU(2)_L$ Higgs doublet of SM is extended to a 
$\SU(2)_L \times \SU(2)_R$ bidoublet. The model
naturally accommodates the discrete parity violation of SM as a result of
spontaneous symmetry breaking of $\SU(2)_R \times \U(1)_{B-L}$, and
also elegantly explains small neutrino masses via the seesaw mechanism.
The model adds new heavy Higgs $\SU(2)_L$ and $\SU(2)_R$ triplets,
henceforth called $\Delta_L$, $\Delta_R$, in order
to implement the seesaw mechanism.
The spontaneous breaking of $Z_2$-discrete symmetry $P$ together with the 
gauge group $\SU(2)_R \times \U(1)_{B-L}$ is achieved by providing
a vacuum expectation value (vev) to $\Delta_R$. This is one of the
most important examples of models with $Z_2$-symmetry breaking.

An interesting variation of LRSM  was provided by
Chang, Mohapatra and Parida \cite{Parida} who proposed a 
`parity decoupled' version  (PLRSM) of the left right symmetric
Standard Model. The gauge symmetry of this model is the same as that
of LRSM.
However as shown by the authors, in this model it is possible to
break the discrete $Z_2$-symmetry $P$ at a higher scale without breaking 
the left right symmetric gauge group
$\SU(3)_c \times \SU(2)_L \times \SU(2)_R \times \U(1)_{B-L}$.
This is done by introducing an additional scalar Higgs field $\eta$ 
which is 
singlet under the left right symmetric gauge group, but transforms as
$\eta \mapsto -\eta$ under $P$. In other words, $P$ implements this 
discrete
symmetry as well as space time parity, without affecting the other
particles. At very high energies the
model is parity invariant. At a slightly lower scale $M_P$, $\eta$ takes a 
non-zero vacuum expectation value (vev) which breaks the 
$Z_2$-symmetry $P$ but leaves the
gauge symmetry unaffected. The $SU(2)_R \times U(1)_{B-L}$ gauge symmetry
is broken at an even lower scale $M_R$. This decoupling of parity 
breaking and gauge symmetry breaking scales allows one to have 
right handed $W_R$ and $Z_R$ bosons at a few TeV as well as an intermediate
scale of partial unification into gauge groups like the Pati-Salam
$\SU(4)_c \times \SU(2)_L \times \SU(2)_R$ at scales of a few hundred
TeV \cite{Parida}. Some of these features may show up in upcoming
collider experiments. Other partial unification features have low 
energy implications like
neutron-antineutron oscillations and neutrinoless double beta decay
which may also be detected in the near future with more sensitive 
experiments. These issues make this model experimentally attractive.

Two Higgs doublet models (2HDM) \cite{Branco} are
studied as examples of models `just beyond SM' as they 
provide more possibilities of electroweak
baryogenesis by containing more sources of CP violation and more scope
of first order phase transitions, which is not the case with the usual
SM \cite{ClineSurvey, Cline2HDM}. These models have two 
(or more) $\SU(2)_L$ Higgs doublets, 
conventionally called $\Phi_1$ and $\Phi_2$.
In order to prevent disastrous consequences of flavour changing
neutral currents, the quark and Higgs sectors in 2HDMs are often required
to satisfy a discrete $Z_2$-symmetry like $\Phi_2 \mapsto -\Phi_2$, 
$d^i_R \mapsto -d^i_R$ or some other similar symmetry. 
This scenario is called Type II 2HDM, where the up-type right handed
quarks couple only to $\Phi_1$ and the down-type right handed quarks
couple only to $\Phi_2$. It is the  most popular
2HDM as it arises as the effective low energy theory
for important BSM models like supersymmetry and axion models
\cite{Branco}. As we will see later, it can arise also as the
effective low energy theory of left right symmetric models like
LRSM and PLRSM. 
This Type II $Z_2$-symmetry is broken when a certain linear combination
$-\sin\beta \Phi_1 + e^{-i\theta} \cos\beta \Phi_2$ gets a vev of $v_1$ in
an extended scenario of electroweak symmetry breaking \cite{Branco}.

In the next section, 
we review the status of several $Z_2$-symmetric models where 
cosmological domain walls can arise. 
We also recall the mechanism of Preskill et al \cite{PreskillTrivedi}, 
relying on QCD instanton anomaly effects,
that induce a pressure difference across domain walls arising
from $Z_2$-symmetry breaking in Type II 2HDM.
Following this in Sec. \ref{sec:PLRSM}, we perform a detailed analysis 
of high scale 
domain wall creation in PLRSM, incorporating the best known
bounds on the parameter space
of LRSM. In the next section \ref{sec:TwoHDM} we show how the 
Type II 2HDM arises as an effective
low energy theory in the LRSM and PLRSM models. 
Section~\ref{sec:DWformation} describes the two types of domain walls 
formed one after the other due to two stages of $Z_2$-symmetry breaking
in PLRSM and LRSM.
In Sec. \ref{sec:DWremoval} we show that both types domain walls that are 
formed can be successfully removed at almost the same time due to 
an energy bias caused by the Preskill et al. result and random fluctuations
in the numbers of certain types of Type II 2HDM subdomains inside an
LRSM/PLRSM domain.
The removal occurs before Big Bang Nucleosynthesis (BBN) 
at temperatures of around $2$ GeV, and does not require any
$Z_2$-symmetry breaking non-renormalisable effects.
Annihilation of domain walls gives rise to gravitational waves. 
In Sec. \ref{sec:GWsignals}
we show that the peak frequency of these waves arising in our mechanism
lies in the 
$10^{-7}$--$10^{-6}~\mbox{Hz}$ band, corresponding to annhilation 
temperatures of $1$--$10$ GeV.
These frequencies are detectable by
existing and proposed pulsar timing array
based gravitational wave detectors like NANOGrav, SKA etc. 
\cite{SKA,NANOGRAV:2023}. The
recent NANOGrav results \cite{NANOGRAV:2023} rule out our DW 
collapse model for higher
values of parity breaking scale above $10^7$ GeV. Our DW collapse model
with parity breaking scales below $10^7$ GeV  remains consistent with 
the current NANOGrav results and has a good
chance of being seriously tested in future pulsar timing based experiments.
We finally conclude with the implications of
this study in Section~\ref{sec:conclusion}.

\section{Cosmology with intermediate scale domain walls}
\label{sec:cosmoDW}
In this paper we study cosmological implications of the breaking
of discrete $Z_2$-symmetry in BSM models like the ones discussed above.
Breaking of $Z_2$-symmetry gives
rise to a network of domain walls in the early universe separating 
domains which can be one of two types
\cite{Kibble:1980, KLS:1982, HK:1995}. For  example, breaking of the
$\eta \mapsto -\eta$ discrete parity symmetry in PLRSM model at energy
scale $M_P$ gives rise to domain walls separating two types of domains.
In one type of domain the $\eta$ Higgs field takes the
vev $M_P$ and in the other, the vev $-M_P$. We shall call the first
type of domain as `positive-high' and the second type as
`negative-high'.
The term `high' is there to remind us that
we are at a high energy scale of $Z_2$-symmetry breaking viz. at $M_P$.
This feature of PLRSM forming its own domain wall has not 
been noticed before, though
in their original paper
Chang, Mohapatra and Parida discussed how certain symmetry breaking chains
descending from $\SO(10)$ to SM avoid $\SO(10)$ domain wall formation
whereas other symmetry breaking chains do not.

In the usual LRSM model,
the breaking of $\SU(2)_L \leftrightarrow \SU(2)_R$ left right
discrete symmetry occurs at an energy scale $M_R$. In one type of
domain the triplet Higgs $\Delta_R$ takes vev $M_R$ leading to 
our familiar left handed SM vacuum.
In the other type of domain $\Delta_L$ takes vev $M_R$ leading to a
right handed SM vacuum
where $\SU(2)_L$ breaks and $\SU(2)_R \times \U(1)_Y$ controls the
electroweak interactions. This is another example of positive-high and
negative-high domain where the term `high' reminds us that the scale
of $Z_2$-symmetry breaking is at high energy of $M_R$.

In Type II 2HDM models, during 
$Z_2$-symmetry breaking 
$-\sin\beta \Phi_1 + e^{-i\theta} \cos\beta \Phi_2$ gets a 
vev of $v_1$ in one type
of domain and a different vev of $v'_1$ in the other type. Henceforth,
we shall call the first type of domain as `positive-low' and the second
type as `negative-low'.
The term `low' is there to remind us that
we are at a lower energy scale of $Z_2$-symmetry breaking $M_{\THDM}$, 
near the electroweak scale.

In the absence of other factors, these domain
walls will be sufficiently long lived so as to conflict with standard
cosmology. Even if the $Z_2$-symmetry is only 
approximate, it is still
possible in some scenarios to end up with long lived domain walls
\cite{Gelmini1989CosmologyBreaking}.
The usual way to remove these domain walls 
is to introduce explicit terms softly breaking the discrete $Z_2$
symmetry to the Lagrangian. For example, Planck scale 
suppressed non-renormalizable operators  were added in
\cite{LR:1993, RS:1994} to the Lagrangian
leading to instability to the domain walls. However this introduction of 
Planck suppressed operators may not work for all
gauge symmetric models. 
In the Supersymmetric Left-Right symmetric Models (SUSYLR) or ultraviolet
completions theoreof, with all Higgs 
carrying gauge charges, it is possible to introduce Planck scale 
suppressed terms or soft SUSY breaking terms that are well regulated. 
One can then demand that 
the new operators ensure sufficient pressure across the domain walls 
that the latter disappear before Big Bang Nucleosynthesis (BBN). 
This requirement has been discussed in detail in 
\cite{Mishra:2010,Aulakh:1998,Borah:2011,SO10DomainWall}.
It has also been shown that domain walls in SUSYLR can be successfully
exploited for baryogenesis via leptogenesis while remaining consistent 
with low
energy observables like the electric dipole moment of the electron,
with a gravitational wave signature that can be detected in upcoming
experiments \cite{Banerjee:leptogenesis}. 
More recent works have studied the
gravitational wave consequences of domain wall collapse 
\cite{Borah:2022wdy,Borboruah:2022eex}  
as well as bubble wall collapse
\cite{Borboruah:2022eex,Brdar:2019fur,Li:2020eun} 
in LRSM, where the domain wall collapse is initiated by adding 
tiny explicit terms breaking the discrete left-right symmetry. In a
related vein, there has been work \cite{Dror:2019syi}
studying the gravitational wave consequences of models
providing a seesaw mechanism together with thermal leptogenesis.

Domain wall formation due to $Z_2$-symmetry breaking in 2HDM models 
was studied recently by Chen et al. \cite{Chen}. The authors showed
that tiny explicit CP violating terms in the 2HDM Higgs 
potential can remove
the domain walls successfully before BBN, and studied the gravitational 
wave and electron EDM consequences of the consequent domain wall collapse.

Preskill et al. \cite{PreskillTrivedi} have investigated domain wall
removal in Type II 2HDM by showing that the $Z_2$-symmetry in Type II 2HDM 
is anomalous. 
More precisely, they point out that the instanton vertex in quantum
chromodynamics (QCD) makes 
the two types of domains viz. positive-low and negative-low 
non-degenerate. In other words one of the types
of domains,
say positive-low, has a slightly lower energy density than the other type
which cannot be seen in the classical perturbative theory but only
in the quantum non-perturbative theory.
Preskill et al. \cite{PreskillTrivedi}  relate the QCD instanton 
vertex to the QCD $\theta$ angle
and provide  a temperature dependent expression for the energy difference
between the two types of domains. The energy difference falls
rapidly with temperature which implies that the domain walls can only
be removed after electroweak symmetry breaking, probably at a temperature
not far from the QCD scale of 340 MeV. Their analysis is preliminary but
shows that domain walls in Type II 2HDM without explicit CP 
violation need not be cosmologically catastropic. 

In this work, instead of adding non-renormalisable operators to
the Lagrangian or invoking explicit CP violating terms, we take 
a different approach to domain wall removal. We leverage the QCD 
anomaly based mechanism of Preskill et al. \cite{PreskillTrivedi},
and show that it can
lead to successful $Z_2$-domain wall removal in all the above 
examples of $Z_2$-symmetry breaking, including LRSM and PLRSM
which do not obviously involve the Type II 2HDM, when combined with
statistical properties of random events. In particular, our
method does not require the addition of explicit CP violating terms
to the 2HDM Higgs potential, which was the strategy undertaken by
Chen et al. \cite{Chen}.

Let us consider the PLRSM model as a running example. At the high scale
$M_P$, two types of domains, positive-high and negative-high, are
formed.
At a lower energy scale $M_\THDM$ when the Type II $Z_2$-symmetry
of 2HDM breaks, two new types of subdomains, positive-low and negative-low,
are created.
At this scale a domain of positive-high type will break up further into 
a bunch of positive-low and negative-low type subdomains.
The same will hold for a
domain of negative-high type. Percolation theory describes the 
numbers, sizes and spatial distribution of the low type subdomains within 
a high type domain. 
Whether a particular positive-high domain
gives rise to more positive-low or negative-low subdomains is a random
variable where either outcome occurs with probability very close to $0.5$. 
The theory of random variables tells us that with 
probability nearly $0.25$, a non-trivial majority of positive-high 
domains will have a non-trivial excess of positive-low subdomains, 
and a non-trivial majority of negative-high domains will have a 
non-trivial excess of negative-low subdomains.
Non-trivial majority or non-trivial excess here means that the 
chance excess is around
square root of the total number of a suitable class of domains / 
subdomains.
The QCD anomaly argument of Preskill et al. \cite{PreskillTrivedi} then 
kicks in and shows
that, on average, the positive-high type domains have lower
energy density as compared to negative-high type domains. 
We show that this feature
can give rise to enough pressure difference to remove
both the high type as well as the low type domain walls almost 
simultaneously in the radiation 
era of the universe well before BBN. 
Both types of walls are annihilated at temperatures around 2 GeV
for parity breaking scales ranging from $10^6$ to $10^7$ GeV.
Previous works like 
\cite{Coulson1996BiasedWalls} have considered probabilistic percolation
based arguments to show the decay of domain walls and eventual domination
of one type of domain in the early universe. However they assume
the presence of
a primordial probabilistic bias between the two types of domains.
The present paper differs from them in assuming no bias between the
types of domains at the time of their formation. Rather it shows that 
natural probabilistic fluctuations in
the number of domains of each type combined with a QCD anomaly that 
manifests itself at a much lower temperature suffice to remove all domain
walls effectively in the early universe. The conclusion is that
explicit soft discrete symmetry breaking terms are not the only viable
way to successfully remove domain walls; pure statistical fluctuations 
combined with an inevitable QCD anomaly can also remove them.
In other words, it is fully possible that the universe we observe 
is a purely chance  survival of one out of the four distinct possibilities.
Our result is agnostic with respect to this provenance.

\section{The PLRSM model}
\label{sec:PLRSM}
The gauge group of PLRSM is the left-right symmetric group
$\SU(3)_c \times \SU(2)_L \times \SU(2)_R \times \U(1)_{B-L}$.
The Higgs sector consists of an $\SU(2)_L \times \SU(2)_R$ Higgs
bidoublet $\Phi$, an $\SU(2)_L$ Higgs triplet $\Delta_L$,
an $\SU(2)_R$ Higgs triplet $\Delta_R$, and a Higgs singlet $\eta$.
All the Higgs fields are singlets under the $\SU(3)_c$ colour group.
Under the left right electroweak gauge group, the Higgs fields transform
as follows:
\[
\Phi \stackrel{(U_L,U_R,e^{i\theta})}{\mapsto} 
U_L \Phi U_R^\dag, ~~
\Delta_L \stackrel{(U_L,U_R,e^{i\theta})}{\mapsto} 
e^{i\theta} U_L \Delta_L U_L^\dag, ~~
\Delta_R \stackrel{(U_L,U_R,e^{i\theta})}{\mapsto} 
e^{i\theta} U_R \Delta_R U_R^\dag, ~~
\eta \stackrel{(U_L,U_R,e^{i\theta})}{\mapsto} 
\eta.
\]
Note that $\tPhi = \tau_2 \Phi^* \tau_2$ transforms in same manner as
$\Phi$ under the gauge action. 
Under discrete parity, these Higgs fields transform as follows:
\[
\Delta_L \leftrightarrow \Delta_R, ~~~
\Phi \leftrightarrow \Phi^\dag, ~~~
\eta \leftrightarrow -\eta.
\]
The electric charge is given by
$
Q = T_{3L} + T_{3R} + \frac{B-L}{2}.
$
The charge zero condition forces the vevs of the Higgs fields to be
\begin{equation}
\label{eq:HiggsVevs}
\vev{\Phi}  =  
\left(
\begin{array}{c c}
\kappa_1 &        0                 \\
     0   &  \kappa_2 e^{i\theta_2} 
\end{array}
\right), ~~ 
\vev{\Delta_L}  =  
\left(
\begin{array}{c c}
0                 & 0 \\
d_L e^{i\theta_L} & 0 
\end{array}
\right), ~~
\vev{\Delta_R}  =  
\left(
\begin{array}{c c}
0   & 0    \\
d_R & 0 
\end{array}
\right),
\end{equation}
where the placement of complex phases is the customary one obtained
by gauge invariance and the real numbers are all positive
\cite{BhupalDev}.

We let $g_L$, $g_R$ and $g_{B-L}$ denote the couplings of the gauge
groups $\SU(2)_L$, $\SU(2)_R$ and $\U(1)_{B-L}$ respectively.
Their values can ultimately be
related to the SM couplings $g_2$ and $g_Y$ of the $\SU(2)_L$ and
$\U(1)_Y$ gauge groups.
At very high energies, the PLRSM model satisfies discrete parity 
symmetry in
addition to gauge symmetry which implies that $g_L = g_R$ at very high
energies.
The renormalisable Higgs potential can 
be written as
$
V_r = V_\eta + V_\Phi + V_\Delta + V_{\Delta \Phi} + 
      V_{\eta \Phi} + V_{\eta \Delta},
$
where  the individual terms are the most general ones allowed by 
gauge and parity symmetry \cite{Parida, BhupalDev} viz.
\begin{equation}
\label{eq:Veta}
V_\eta = -\mu_\eta^2 \eta^2 + \delta_1 \eta^4,
\end{equation}
\begin{equation}
\label{eq:VetaPhi}
V_{\eta \Phi} = 
\gamma_1 \eta^2 \Tr \Phi^\dag \Phi +
\gamma_2 \eta^2 (\Tr \Phi^\dag \tPhi + \Tr \tPhi^\dag \Phi),
\end{equation}
\begin{equation}
\label{eq:VetaDelta}
V_{\eta \Delta} = 
M \eta (\Tr \Delta_L^\dag \Delta_L - \Tr \Delta_R^\dag \Delta_R) +
\delta_2 \eta^2 (\Tr \Delta_L^\dag \Delta_L + \Tr \Delta_R^\dag \Delta_R).
\end{equation}
\begin{equation}
\label{eq:VPhi}
\begin{array}{rcl}
V_\Phi 
& = &
-\mu_\Phi^2 \Tr \Phi^\dag \Phi -
\mu_{\tPhi}^2 (\Tr \Phi^\dag \tPhi + \Tr \tPhi^\dag \Phi) \\
&   &
{} +
\lambda_1 (\Tr \Phi^\dag \Phi)^2 +
\lambda_2 ((\Tr \Phi^\dag \tPhi)^2 + (\Tr \tPhi^\dag \Phi)^2)  \\
&    &
{} +
\lambda_3 \Tr \tPhi \Phi^\dag \Tr \tPhi^\dag \Phi +
\lambda_4 \Tr \Phi^\dag \Phi (\Tr \Phi^\dag \tPhi + \Tr \tPhi^\dag \Phi) \\
&    &
{} +
\lambda_5 \Tr \Phi^\dag \Phi \Phi^\dag \Phi +
\lambda_6 (\Tr \tPhi^\dag \Phi \tPhi^\dag \Phi +
	   \Tr \Phi^\dag \tPhi \Phi^\dag \tPhi) \\
&    &
{} +
\lambda_7 (\Tr \Phi^\dag \Phi \Phi^\dag \tPhi +
	   \Tr \tPhi^\dag \Phi \Phi^\dag \Phi) +
\lambda_8 (\Tr \Phi^\dag \Phi \tPhi^\dag \tPhi +
	   \Tr \tPhi^\dag \Phi \Phi^\dag \tPhi), 
\end{array}
\end{equation}
\begin{equation}
\label{eq:VDelta}
\begin{array}{rcl}
V_\Delta 
& = &
\mu_\Delta^2 (\Tr \Delta_L^\dag \Delta_L + \Tr \Delta_R^\dag \Delta_R) \\
&    &
{} +
\rho_1 ((\Tr \Delta_L^\dag \Delta_L)^2 +
	(\Tr \Delta_R^\dag \Delta_R)^2
       ) \\
&    &
{} +
\rho_2 (\Tr \Delta_L \Delta_L \Tr \Delta_L^\dag \Delta_L^\dag +
	\Tr \Delta_R \Delta_R \Tr \Delta_R^\dag \Delta_R^\dag
       ) \\
&    &
{} +
\rho_3 \Tr \Delta_L^\dag \Delta_L \Tr \Delta_R^\dag \Delta_R +
\rho_4 (\Tr \Delta_L \Delta_L \Tr \Delta_R^\dag \Delta_R^\dag +
	\Tr \Delta_L^\dag \Delta_L^\dag \Tr \Delta_R \Delta_R
       ) \\ 
&    &
{} +
\rho_5 (\Tr \Delta_L^\dag \Delta_L \Delta_L^\dag \Delta_L +
        \Tr \Delta_R^\dag \Delta_R \Delta_R^\dag \Delta_R
       ) \\
&    &
{} + 
\rho_6 (\Tr \Delta_L^\dag \Delta_L^\dag \Delta_L \Delta_L +
        \Tr \Delta_R^\dag \Delta_R^\dag \Delta_R \Delta_R
       ),
\end{array}
\end{equation}
\begin{equation}
\label{eq:VDeltaPhi}
\begin{array}{rcl}
V_{\Delta \Phi}
& = &
\alpha_1 \Tr \Phi^\dag \Phi 
(\Tr \Delta_L^\dag \Delta_L + \Tr \Delta_R^\dag \Delta_R) \\ 
&   &
{} +
\alpha_2 (\Tr \Delta_L \Delta_L^\dag \Tr \tPhi \Phi^\dag +
          \Tr \Delta_R \Delta_R^\dag \Tr \tPhi^\dag \Phi
	 ) \\
&   &
{} +
\alpha_2^* (\Tr \Delta_L \Delta_L^\dag \Tr \tPhi^\dag \Phi +
            \Tr \Delta_R \Delta_R^\dag \Tr \tPhi \Phi^\dag
           ) \\
&   &
{} +
\alpha_3 
(\Tr \Phi \Phi^\dag \Delta_L \Delta_L^\dag + 
 \Tr \Phi^\dag \Phi \Delta_R \Delta_R^\dag
) +
\alpha_4 
(\Tr \tPhi \tPhi^\dag \Delta_L \Delta_L^\dag + 
 \Tr \tPhi^\dag \tPhi \Delta_R \Delta_R^\dag
) \\
&   &
{} +
\alpha_5 (\Tr \tPhi \Phi^\dag \Delta_L \Delta_L^\dag + 
          \Tr \tPhi^\dag \Phi \Delta_R \Delta_R^\dag 
         ) +
\alpha_5^* (\Tr \Phi \tPhi^\dag \Delta_L \Delta_L^\dag +
            \Tr \Phi^\dag \tPhi \Delta_R \Delta_R^\dag 
           ) \\
&    &
{} +
\beta_1 (\Tr \Phi \Delta_R \Phi^\dag \Delta_L^\dag +
         \Tr \Phi^\dag \Delta_L \Phi \Delta_R^\dag
        ) +
\beta_2 (\Tr \tPhi \Delta_R \Phi^\dag \Delta_L^\dag +
         \Tr \tPhi^\dag \Delta_L \Phi \Delta_R^\dag
        ) \\
&    &
{} +
\beta_3 (\Tr \Phi \Delta_R \tPhi^\dag \Delta_L^\dag +
         \Tr \Phi^\dag \Delta_L \tPhi \Delta_R^\dag
        ) +
\beta_4 (\Tr \tPhi \Delta_R \tPhi^\dag \Delta_L^\dag +
         \Tr \tPhi^\dag \Delta_L \tPhi \Delta_R^\dag
        ),
\end{array}
\end{equation}
We assume that $\mu_\eta$ is much larger than the other masses in 
the model. So to a very good approximation we can minimise the total
potential by first minimising the term
$V_\eta$ followed by sequentially minimising the other terms. 
Minimising $V_\eta$ in Equation~\ref{eq:Veta}
leads to the conclusion that at a high scale 
$
M_P \approx \frac{\mu_\eta}{\sqrt{2 \delta_1}}
$
the $\eta$ field takes the vev $\vev{\eta} = M_P$. This gives
$
\vev{V_\eta} = -\frac{\mu_\eta^4}{4 \delta_1} =
-\delta_1 M_P^4.
$
Using Equations~\ref{eq:VDelta}, \ref{eq:VetaDelta}, the effective
mass terms for $\Delta_L$ and $\Delta_R$ become
\begin{equation}
\label{eq:muDeltaLR1}
\mu_{\Delta_R}^2 = 
\mu_\Delta^2 - M \cdot M_P + \delta_2 M_P^2, ~~~
\mu_{\Delta_L}^2 = 
\mu_\Delta^2 + M \cdot M_P + \delta_2 M_P^2.
\end{equation}
Restricting ourselves to only those terms that survive the setting of 
the bidoublet vevs to zero, we get
\begin{equation}
\label{eq:VDeltaVev}
\begin{array}{r c l}
\vev{V_{\Delta}} + \vev{V_{\eta \Delta}}
& = &
\mu_{\Delta_L}^2 \Tr \Delta_L^\dag \Delta_L +
\mu_{\Delta_R}^2 \Tr \Delta_R^\dag \Delta_R \\
&   &
{} +
\rho_1 ((\Tr \Delta_L^\dag \Delta_L)^2 +
	(\Tr \Delta_R^\dag \Delta_R)^2
       ) +
\rho_3 \Tr \Delta_L^\dag \Delta_L \Tr \Delta_R^\dag \Delta_R \\
&    &
{} +
\rho_5 (\Tr \Delta_L^\dag \Delta_L \Delta_L^\dag \Delta_L +
        \Tr \Delta_R^\dag \Delta_R \Delta_R^\dag \Delta_R
       ).
\end{array}
\end{equation}

The triplet Higgs field $\Delta_R$ gets a non-zero vev at a 
lower scale $M_R$ breaking down $\SU(2)_R \times \U(1)_{B-L}$ to 
$\U(1)_Y$, $\Phi$ gets a non-zero vev at an even
lower scale breaking down $\SU(2)_L \times \U(1)_Y$ to 
$\U(1)_{\mathrm{em}}$. This decouples the parity breaking scale $M_P$ from
the gauge breaking scale $M_R$. Note that $\mu_{\Delta_L}^2 > 0$ 
as a result
of which $\Delta_L$ gets a zero or almost zero vev. The hierarchy
achieved by the fine tuning allows us to conclude with a high degree
of precision that 
$M_R \approx \frac{\mu_{\Delta_R}}{\sqrt{2(\rho_1 + \rho_5)}}$.
In fact after parity breaking at scale $M_P$, the $\Delta_L$ fields
end up with higher masses than the $\Delta_R$ fields even before gauge
symmetry breaking at the lower scale $M_R$. A consequence of this is that 
the gauge couplings $g_L$ and $g_R$ are no longer equal between the energy
$M_P$ and $M_R$ owing to their different runnings under their respective 
renormalisation group equations.
All these features are achieved by implementing the fine 
tuning condition $\mu_{\Delta_R}^2 < 0$, 
$
\mu_\Phi^2 \approx M_W^2 
\ll M_R^2 \approx |\mu_{\Delta_R}^2| \ll \delta_2 M_P^2 < M_P^2.
$   

Notice now that $V_\eta$ has another minimum at
$\vev{\eta} = -M_P$. This new minimum cannot be gauged away as the
gauge has been fixed by the customary electromagnetic charge zero 
convention of the vevs of $\Delta_L$, $\Delta_R$ and $\Phi$. This new
minimum leads to a different vev assignment to the other Higgs fields
because, using Equations~\ref{eq:VDelta}, \ref{eq:VetaDelta},
the effective mass terms for $\Delta_L$ and $\Delta_R$ now become
\begin{equation}
\label{eq:muDeltaLR2}
\mu_{\Delta_L}^2 = 
\mu_\Delta^2 - M \cdot M_P + \delta_2 M_P^2, ~~~
\mu_{\Delta_R}^2 = 
\mu_\Delta^2 + M \cdot M_P + \delta_2 M_P^2.
\end{equation}
In this scenario, $\Delta_L$ gets a non-zero vev 
at the lower scale $M_R$ breaking down $\SU(2)_L \times \U(1)_{B-L}$ to 
$\U(1)_Y$, $\Phi$ gets a non-zero vev at an even
lower scale breaking down $\SU(2)_R \times \U(1)_Y$ to 
$\U(1)_{\mathrm{em}}$. This scenario is the mirror image of the standard
scenario and can be thought of as the right handed Standard Model.

From a cosmological viewpoint, the early universe breaks down at the
high energy scale of $M_P$ into a network of domains of two types.
In the positive high domain, $\vev{\eta} = M_P$. Later on at a lower energy
scale $M_R$, the `right handed triplet Higgs fields' take 
non-zero vevs and so the positive high domain eventually leads to
the familiar left handed SM domain.
In the negative high domain, $\vev{\eta} = -M_P$. Later on at a lower 
energy scale $M_R$, the `left handed triplet Higgs fields' take 
non-zero vevs and so the negative high domain eventually leads to
the right handed SM domain. 
To summarise, the two sets of vevs are
\begin{equation}
\label{eq:domainvevs}
\begin{array}{r c l}
(\vev{\eta}, \vev{d_L}, \vev{d_R}) 
& = &
(M_P, 0, M_R), 
~~~
\mbox{Positive-high, LH}, \\
(\vev{\eta}, \vev{d_L}, \vev{d_R}) 
& = &
(-M_P, M_R, 0), 
~~~
\mbox{Negative-high, RH}, \\
M_P & \approx & \frac{\mu_\eta}{\sqrt{2 \delta_1}}, \\
M_R & \approx &
\frac{\sqrt{|\mu_\Delta^2 - M \cdot M_P + \delta_2 M_P^2|}
     }{\sqrt{2 (\rho_1 + \rho_5)}} 
\;\ll\; M_P.
\end{array}
\end{equation}
Since at the left right gauge symmetry breaking scale $M_R$ the 
bidoublet has
zero vev, it has not been mentioned in the above equation.

The breaking of the discrete parity symmetry and the formation of the 
two types of high energy domains also leads to a topological domain wall 
between them. The presence
of this wall or energy barrier conflicts with current cosmology. Several
earlier works have discussed how a similar wall that occurs in the
minimal LRSM or its supersymmetric variant may be made to disappear
fast enough so as to be consistent with present day observations. All
these works either add $Z_2$-symmetry breaking
Planck suppressed non-renormalisable operators 
\cite{RS:1994,Mishra:2010}
or soft SUSY breaking terms 
\cite{Borah:2011}
to the Lagrangian in order to remove the
domain walls.
However the formation of domain wall in the 
PLRSM has not
been noticed before. Analogous to domain wall removal in the minimal
LRSM, it is possible to add $Z_2$-symmetry breaking
Planck suppressed non-renormalisable operators to the Lagrangian of
PLRSM in order to remove its domain walls. However in this paper we
shall not do so. Instead, we shall leverage the QCD anomaly 
based mechanism of Preskill et al. \cite{PreskillTrivedi} and combine
it with statistical arguments in order to ensure successful domain wall
removal. The details follow in the subsequent sections.

The best model independent lower bounds on the energy scale $M_R$ of 
left right gauge symmetry breaking are of the order of around 4.7 TeV
and come from the ATLAS experiment \cite{AtlasWR:2021}. 
Since in the PLRSM model the scale
$M_P$ of discrete parity breaking is assumed to be decoupled and
significantly
higher than $M_R$, we shall take a lower bound of
$M_P > 10^6\,\GeV$ in our work.

\section{Type II 2HDM as an effective low energy theory}
\label{sec:TwoHDM}
In this section we shall see how the Type II 2HDM arises as an effective
low energy theory in the LRSM \cite{MohapatraLR} and PLRSM \cite{Parida}
models.

In both PLRSM and LRSM, the Yukawa part of the Lagrangian involving
the bidoublet $\Phi$, neglecting generation indices, is of the 
following form
\begin{equation}
\label{eq:Yukawa}
\mathcal{L}_{\mathrm{Yukawa}} =
h_1 \bar{q}_L 
\Phi 
q_R +
h_2 \bar{q}_L 
\tPhi 
q_R +
h_3 \bar{l}_L 
\Phi 
l_R +
h_4 \bar{l}_L 
\tPhi 
l_R,
\end{equation}
where the quarks and leptons are organised into doublets under the
action of $\SU(2)_L$ or $\SU(2)_R$ as appropriate i.e.
\begin{equation}
\label{eq:doublets}
\hspace*{-23mm}
\Phi = 
\left(
\begin{array}{c c}
\phi_{11} & \phi_{12} \\
\phi_{21} & \phi_{22}
\end{array}
\right),
\tPhi = 
\left(
\begin{array}{c c}
\phi_{22}^* & -\phi_{21}^* \\
-\phi_{12}^* & \phi_{11}^*
\end{array}
\right),
q_L = 
\left(
\begin{array}{c}
u_L \\
d_L
\end{array}
\right),
q_R = 
\left(
\begin{array}{c}
u_R \\
d_R
\end{array}
\right),
l_L = 
\left(
\begin{array}{c}
\nu_L \\
e_L
\end{array}
\right),
l_R = 
\left(
\begin{array}{c}
\nu_R \\
e_R
\end{array}
\right)\!\!.
\end{equation}
After the breaking of the $\SU(2)_L \times \SU(2)_R$ gauge symmetry
at scale $M_R$, the above LRSM Yukawa terms become 2HDM Yukawa terms.
In the LH or positive-high domain, the resulting 2HDM Yukawa terms become
\begin{equation}
\label{eq:Yukawa2HDMLH}
\begin{array}{rcl}
\mathcal{L}_{\mathrm{2HDMLH}} 
& = &
\bar{q}_L 
\left(
\begin{array}{c}
h_1 \phi_{11} + h_2 \phi_{22}^* \\
h_1 \phi_{21} - h_2 \phi_{12}^* 
\end{array}
\right) u_R +
\bar{q}_L 
\left(
\begin{array}{c}
h_1 \phi_{12} - h_2 \phi_{21}^* \\
h_1 \phi_{22} + h_2 \phi_{11}^*
\end{array}
\right) d_R \\
& &
{} +
\bar{l}_L 
\left(
\begin{array}{c}
h_3 \phi_{11} + h_4 \phi_{22}^* \\
h_3 \phi_{21} - h_4 \phi_{12}^* 
\end{array}
\right) \nu_R +
\bar{l}_L 
\left(
\begin{array}{c}
h_3 \phi_{12} - h_4 \phi_{21}^* \\
h_3 \phi_{22} + h_4 \phi_{11}^*
\end{array}
\right) e_R, 
\end{array}
\end{equation}
under the gauge group $\SU(2)_L \times U(1)_Y$. Ignoring the high mass
right handed neutrino $\nu_R$, assuming $h_3$, $h_4$ to be proportional to
$h_1$, $h_2$ for simplicity, the Yukawa Lagrangian in the LH domain becomes
the standard Type II 2HDM Lagrangian 
\begin{equation}
\label{eq:Yukawa2HDMLHTypeII}
\mathcal{L}_{\mathrm{2HDMLH}} = 
\sqrt{h_1^2 + h_2^2} \, \bar{q}_L \Phi'_1 u_R +
\sqrt{h_1^2 + h_2^2} \, \bar{q}_L \Phi'_2 d_R +
\frac{h_3}{h_1} \sqrt{h_1^2 + h_2^2} \, \bar{l}_L \Phi'_2 e_R
\end{equation}
for the two $\SU(2)_L$ Higgs doublets
\begin{equation}
\label{eq:2HDMLHDoublets}
\Phi'_1 =
\frac{1}{\sqrt{h_1^2 + h_2^2}}
\left(
\begin{array}{c}
h_1 \phi_{11} + h_2 \phi_{22}^* \\
h_1 \phi_{21} - h_2 \phi_{12}^* 
\end{array}
\right),
\Phi'_2 =
\frac{1}{\sqrt{h_1^2 + h_2^2}}
\left(
\begin{array}{c}
h_1 \phi_{12} - h_2 \phi_{21}^* \\
h_1 \phi_{22} + h_2 \phi_{11}^* 
\end{array}
\right).
\end{equation}
It can be checked that the two doublets above have weak hypercharge
$Y = 1$, and also that they are orthogonal because the constituent
fields $\phi_{ij}$ are distinct, as expected from  a left handed 2HDM.

In the RH or negative-high domain, the resulting 2HDM Yukawa terms become
\begin{equation}
\label{eq:Yukawa2HDMRH}
\hspace*{-10mm}
\begin{array}{rcl}
\lefteqn{
\mathcal{L}_{\mathrm{2HDMRH}} 
} \\
& = &
\bar{u}_L 
\left(
h_1 \phi_{11} + h_2 \phi_{22}^* \;\;\;
h_1 \phi_{12} - h_2 \phi_{21}^* 
\right) q_R +
\bar{d}_L 
\left(
\begin{array}{c}
h_1 \phi_{21} - h_2 \phi_{12}^* \;\;\;
h_1 \phi_{22} + h_2 \phi_{11}^*
\end{array}
\right) q_R \\
& &
{} +
\bar{\nu}_L 
\left(
\begin{array}{c}
h_3 \phi_{11} + h_4 \phi_{22}^* \;\;\;
h_3 \phi_{21} - h_4 \phi_{12}^* 
\end{array}
\right) l_R +
\bar{e}_L 
\left(
\begin{array}{c}
h_3 \phi_{12} - h_4 \phi_{21}^* \;\;\;
h_3 \phi_{22} + h_4 \phi_{11}^*
\end{array}
\right) l_R, 
\end{array}
\end{equation}
under the gauge group $\SU(2)_R \times U(1)_Y$. Ignoring the high mass
left handed neutrino $\nu_L$, assuming $h_3$, $h_4$ to be proportional to
$h_1$, $h_2$ for simplicity, the Yukawa Lagrangian in the RH domain becomes
the standard Type II 2HDM Lagrangian 
\begin{equation}
\label{eq:Yukawa2HDMRHTypeII}
\mathcal{L}_{\mathrm{2HDMLH}} = 
\sqrt{h_1^2 + h_2^2} \, \bar{q}_R \Phi''_1 q_L +
\sqrt{h_1^2 + h_2^2} \, \bar{q}_R \Phi''_2 d_L +
\frac{h_3}{h_1} \sqrt{h_1^2 + h_2^2} \, \bar{l}_R \Phi''_2 e_L
\end{equation}
for the two $\SU(2)_R$ Higgs doublets
\begin{equation}
\label{eq:2HDMRHDoublets}
\Phi''_1 =
\frac{1}{\sqrt{h_1^2 + h_2^2}}
\left(
\begin{array}{c}
h_1 \phi_{11}^* + h_2 \phi_{22} \\
h_1 \phi_{12}^* - h_2 \phi_{21} 
\end{array}
\right),
\Phi''_2 =
\frac{1}{\sqrt{h_1^2 + h_2^2}}
\left(
\begin{array}{c}
h_1 \phi_{21}^* - h_2 \phi_{12} \\
h_1 \phi_{22}^* + h_2 \phi_{11} 
\end{array}
\right).
\end{equation}
Again, the two doublets above have weak hypercharge
$Y = 1$ and are orthogonal, as expected from  a right handed 2HDM.

Thus, to summarise both PLRSM and LRSM reduce to Type II 2HDM in both
LH and RH domains after the $\SU(2)_L \times \SU(2)_R$ gauge symmetry
breaking, provided $h_3$, $h_4$ are proportional to $h_1$, $h_2$
respectively. Without this proportionality condition however, they
do not reduce to 2HDM.

Finally, we would like to discuss the issue of flavour changing
neutral current (FCNC) effects in LRSM, and how they do not apply to
the low energy effective theory of Type II 2HDM that we get as described
above when the proportionality condition on the Yukawa couplings holds.
FCNC effects occur in PLRSM and LRSM after 
the LR symmetry  is broken only if the two resulting $\SU(2)_L$ Higgs 
doublets get vevs 
of the form $\langle \Phi_1 \rangle = (v_1, 0)$,
$\langle \Phi_2 \rangle = (0, v_2 e^{i\theta})$ i.e. the vevs are of
the bidoublet vev form. This can happen if the LR breaking scale is not
too distant from the scale of the Higgs doublet vevs. 
This was discussed in the paper by Zhang et al. \cite{ZhangLRSM:2008}.
However in this 
paper we are considering the scale of Higgs doublet vevs close to the 
electroweak scale and well separated from LR breaking scale. That is why
in our scenario, after LR symmetry breaking, the effective theory becomes
a Type II 2HDM under the proportionality condition. Hence in our model, 
the two Higgs doublets take vevs of the form
$\langle \Phi_1 \rangle = (0, v_1)$,
$\langle \Phi_2 \rangle = (0, v_2 e^{i\theta})$ and there are no flavour
changing effects at tree level. Thus,
our conclusions in this section are consistent
with the observations in Section 4.2 of Zhang et al..

\section{Hierarchical domain wall formation in PLRSM}
\label{sec:DWformation}
In most examples of $Z_2$-domain walls formed at high enough energy 
scales, the 
pressure difference $\delta V$ between the two types of domains is much
smaller than the average energy density $V_0$ of a domain. For example,
in the PLRSM model
because of our fine tuning, the largest contribution to the Higgs 
potential is from $\vev{V_\eta}$. 
Thus the height $V_{0,\hi}$ of the potential barrier occurring at 
$\eta=0$ is
$ -\vev{V_\eta}$, or, 
$V_{0,\hi} \approx |\vev{V_{\eta}}| \approx \delta_1 M_P^4 \approx M_P^4$. 
Hence we can accurately describe
the surface energy density of the high type of domain wall by
$\sigma_\hi \approx M_P^3 \sqrt{\delta_1} \approx M_P^3$ \cite{Saikawa}. 
Because of the $Z_2$-symmetry at high scale, 
$\delta V_\hi(t_{\fo,\hi}) = 0$ at tree level, where $t_{\fo,\hi}$ is
the time of formation of the high type domain walls.
Non-renormalisable effects and quantum corrections are expected to be
quite small compared to $V_{0,\hi}$. Thus, 
$\delta V_\hi(t_{\fo,\hi}) \ll V_{0,\hi}$. In particular
$\frac{\delta V_\hi(t_{\fo,\hi})}{V_{0,\hi}} < 0.795$, as a result of 
which percolation theory
ensures that there are giant domains
of both positive-high and negative-high types formed at scale $M_P$ 
together with large scale domains walls separating them \cite{Saikawa}. 
Moreover, the probabilities $p_{+,\hi}$, $p_{-,\hi}$ of 
forming  positive-high and negative-high type domains are very close
to 0.5 each, since their ratio must satisfy \cite{Saikawa}
\[
\frac{p_{+,\hi}}{p_{-,\hi}} \approx 
\exp\left(-\frac{\delta V_\hi(t_{\fo,\hi})}{V_{0,\hi}}\right) \approx 1. 
\]

Some time after the discrete parity breaking in PLRSM at temperature 
$T_{\fo,\hi} = M_P$ 
resulting in the creation of giant domains of positive-high
and negative-high types, the gauge symmetry
$\SU(2)_L \times \SU(2)_R$ breaks in each domain. 
In the positive-high
domian, $SU(2)_R$ breaks at an energy scale $M_R < M_P$.
In the negative-high
domian, $SU(2)_L$ breaks at energy scale $M_R$. However these are not
discrete symmetry breakings, and so no further subdomains after formed
inside the domains yet.

At this stage, as argued in the previous section, the effective theory 
in the positive-high domain is
the left handed Type II 2HDM with the gauge group
$\SU(3)_C \times \SU(2)_L \times U(1)_Y$, and the effective theory
in the negative-high domain is the right handed Type II 2HDM with
the gauge group $\SU(3)_C \times \SU(2)_R \times U(1)_Y$.
At a still lower temperature $T_{\fo,\lo} = M_\THDM$, the 
Type II 2HDM $Z_2$-symmetry breaks 
resulting in two types of subdomains being formed inside each domain. 
This happens because there are two possible choices for the vevs
of $\Phi_1$, $\Phi_2$ viz.
\begin{eqnarray}
(\vev{\Phi_1}, \vev{\Phi_2}) 
&=& 
(-v_1 \sin\beta, e^{i\theta} v_1 \cos\beta) 
\qquad \textrm{positive-low type}\\
(\vev{\Phi_1}, \vev{\Phi_2}) 
&=& 
(-v_1 \sin\beta, e^{-i\theta} v_1 \cos\beta) 
\qquad \textrm{negative low-type}.
\end{eqnarray}
Another way to express the two choices above is by postulating that
a certain linear combination of $\Phi_1$ and $\Phi_2$ gets two possible
vevs:
\begin{eqnarray}
\vev{-\sin\beta \Phi_1 + e^{-i\theta} \cos\beta \Phi_2} 
&=& 
v_1 \qquad \textrm{positive low-type}
\label{eq:positivelow} \\
\vev{-\sin\beta \Phi_1 + e^{-i\theta} \cos\beta \Phi_2} 
&=& 
v'_1 \qquad \textrm{negative low-type}
\label{eq:negativelow} 
\end{eqnarray}
Note that these ansatz are valid even in the absence of any explicit 
CP violation assumed in the 2HDM potential.

Due to the above spontaneous $Z_2$-symmetry breaking in the Type II 2HDM,
a high type domain breaks up into a bunch of positive-low and 
negative-low subdomains. Let $t_{\fo,\lo}$  be the time
of formation of the low type subdomains.
The effective theory inside all the subdomains formed within a 
positive-high domain is the usual left handed SM.
The effective theory inside all the subdomains formed within a 
negative-high domain is the right handed SM. In this work, we assume
that $M_\THDM \approx 10 M_\EW$ i.e. the 2HDM $Z_2$-symmetry breaking
scale is well separated by around an order of magnitude above 
the electroweak scale.

Preskill, Trivedi, Wilczek and Wise \cite{PreskillTrivedi} show that
for temperatures $T$ above the QCD scale $\Lambda_\QCD = 340~\mathrm{MeV}$ 
\cite{OlivePDG}
the positive-low subdomain, say, has a
lower energy density than the negative low subdomain by an amount
equal to
\begin{equation}
\label{eq:eps}
\delta V_\lo(T) =
8 \times 10^{-4} \, \Lambda_{\QCD} m_u m_d m_s
\left(\frac{\Lambda_\QCD}{\pi T}\right)^8
\left(9 \ln\left(\frac{\pi T}{\Lambda_\QCD}\right)\right)^6,
\end{equation}
where $m_u = 2.16~\mathrm{MeV}$, $m_d = 4.67~\mathrm{MeV}$, 
$m_s = 93.4~\mathrm{MeV}$ \cite{BBNPDG} are the masses of the 
up, down and strange quarks. The energy
difference falls rapidly with temperature. 

The temperature $T_{\fo,\lo}$ at time $t_{\fo,\lo}$ is nothing but
$T_{\fo,\lo} = M_\THDM \approx 10 M_\EW \approx 1000\,\GeV$. 
At this temperature, the energy difference becomes
\[
\delta V_\lo(T_{\fo,\lo}) =
(8 \times 10^{-13}) \cdot 0.34 \cdot 2.16 \cdot 4.67 \cdot 93.4
\left(\frac{0.34}{1000 \pi}\right)^8
\left(9 \ln \left(\frac{1000 \pi}{0.34}\right)\right)^6 < 10^{-33}\,
\GeV^4.
\]
The height $V_{0,\lo}$ of the potential barrier between positive-low 
and negative-low subdomains is around
$M_\EW^4 \approx 10^8\,\GeV^4$ \cite{Chen}. Hence
the probabilities $p_{+,\lo}$, $p_{-,\lo}$ of 
forming  positive-low and negative-low type subdomains are very close
to 0.5 each, since their ratio must satisfy \cite{Saikawa}
\[
\frac{p_{+,\lo}}{p_{-,\lo}} \approx 
\exp\left(-\frac{\delta V_\lo(T_{\fo,\lo})}{V_{0,\lo}}\right) \approx 1. 
\]
Moreover
$\frac{\delta V_\lo(T_{\fo,\lo})}{V_{0,\lo}} < 0.795$; so
percolation theory ensures that there  are giant subdomains
of both positive-low and negative-low types 
inside each high type domain, together with large scale low type
walls separating them \cite{Saikawa}.

\section{Two stage domain wall removal in PLRSM}
\label{sec:DWremoval}
Recall from the previous section that the high type walls are formed
at time $t_{\fo,\hi}$ and temperature $T_{\fo,\hi}$. The surface
tension inside the high type walls is $\sigma_{\hi} \approx M_P^3$, 
where $M_P$ is the scale at which parity is broken. We take 
$T_{\fo,\hi} = M_P$.
Some time later on, the low type walls are formed at time 
$t_{\fo,\lo}$ and temperature $T_{\fo,\lo} \approx 10 M_\EW$. The surface
tension inside the low type walls is $\sigma_{\lo} \approx 
M_\EW^3 \approx 10^6\,\GeV^3$ \cite{Chen}. 

We assume that the high type walls follow the scaling evolution soon after
$t_{\fo,\hi}$  and all the way till $t_{\fo,\lo}$ and even after till
the time of annihilation of high and low type walls.
This is because the high type walls are formed from the
$\eta$ scalar field which only interacts with the singlet $\eta$ 
Higgs particle,
a very massive Higgs. A similar remark would hold for high type walls
in the minimal LRSM where they are formed from the triplet
$\Delta_L$, $\Delta_R$ Higgs particles, which are again massive Higgs.
In particular, the high type walls
do not interact with the 2HDM and SM particles which proliferate inside
both kinds of high type domains at temperatures significantly below
$T_{\fo,\hi}$.
Hence, the radius of curvature of the high type walls at time 
$t_{\fo,\lo}$ is given by the scaling equation \cite{Saikawa}
\begin{equation}
\label{eq:RhiTlo}
R_\hi(t_{\fo,\lo}) = \frac{t_{\fo,\lo}}{A}.
\end{equation}
This is the Hubble radius at time $t_{\fo,\lo}$ divided by a wall parameter
$A = 0.8 \pm 0.1$. 

When low type walls are formed at time $t_{\fo,\lo}$, their correlation
length $\xi_\lo$ is given by \cite{BlasiDWCorrelation}
\begin{equation}
\label{eq:RloTlo}
\xi_\lo = (8 \pi G \sigma_\lo t_{\fo,\lo}^3)^{1/2} = 
\frac{\sigma_\lo^{1/2} t_{\fo,\lo}^{3/2}}{M_\Pl},
\end{equation}
where  $M_{\Pl} = (8 \pi G)^{-1/2} = 2.435 \times 10^{18}\,\GeV$ is
the reduced Planck mass.
The above equation holds because at the time of formation, the surface 
tension in the low type walls has to balance the frictional force that
they encounter due to interaction with the particles in the low type
subdomains moving around.

We now need to deduce some properties of the low type subdomains 
formed within a single high type domain. For concreteness, let us 
concentrate on the positive low subdomains inside a high domain. We
can model the formation of the positive low subdomains via percolation
theory, by fitting a cubic lattice inside a high domain. The low
subdomain correlation length $\xi_\lo$ is taken to be the spacing between 
two nearest
lattice points. The number of lattice points $N_L$ inside a high domain is
now given by
\begin{equation}
\label{eq:Lpoints1}
N_L = \left(\frac{R_\hi(t_{\fo,\lo})}{\xi_\lo}\right)^3 =
\left(\frac{M_\Pl}{A t_{\fo,\lo}^{1/2} \sigma_\lo^{1/2}}\right)^3.
\end{equation}
Since this is occuring in the radiation dominated era of the early
universe, the relation between time and temperature is given by
\cite{Saikawa}
\begin{equation}
\label{eq:tvsT}
T_{\fo,\lo} = 
8.747 \times 10^{-4}
(\frac{g_*(T_{\fo,\lo})}{10})^{-1/4}
t_{\fo,\lo}^{-1/2},
\end{equation}
where $g_*(T_{\fo,\lo})$ is the number of relativistic degrees 
of freedom for the energy density at temperature $T_{\fo,\lo}$.
Plugging this into the expression for the ratio gives
\begin{equation}
\label{eq:Lpoints2}
N_L =
\left(
4.2 \times 10^{3.5}
\frac{M_\Pl (\frac{g_*(T_{\fo,\lo})}{10})^{1/4} T_{\fo,\lo}}
     {A \sigma_\lo^{1/2}}
\right)^3.
\end{equation}

Because two adjacent lattice points are $\xi_\lo$ apart, the choices
of the types of low domains formed at those two points become two
independent random variables with probability $0.5$ of either choice
occuring at a lattice point. In other words, the formation of positive
low subdomains can be treated as a site percolation problem on a simple 
cubic lattice in three dimensions \cite{Stauffer} with $p = 0.5$ 
being the probability that a particular lattice point is occupied.

Maximal sets of adjacent occupied sites are known as clusters.
Percolation theory says that if the occupation probability $p$ is
above the so-called percolation threshold $p_c$, then there is exactly
one giant, aka infinite, cluster that touches all faces of the bounding
cube of the lattice \cite{Stauffer}. For site percolation on the simple
cubic lattice in three dimensions, $p_c = 0.311$ \cite{Stauffer}. 
Since in our case $0.5 = p > p_c = 0.311$, we have exactly one giant
subdomain of positive low type inside a high domain. Similarly we have
exactly one giant subdomain of negative low type inside a high domain.

However, in addition to the unique giant or infinite cluster, there are
many small, aka finite, clusters. Let $P_\infty$ be the probability
that an occupied site belongs to the giant cluster. Then the fraction
$V_f$ of lattice points belonging to finite clusters is given by
$V_f = p - p P_\infty$. According to the scaling laws of percolation 
theory, 
this fraction $V_f$ is
described in terms of the so-called critical exponent $\beta$ 
\cite{Stauffer}:
\begin{equation}
\label{eq:VFinite}
V_f = p - p(p - p_c)^\beta.
\end{equation}
For the simple cubic lattice in three dimensions, $\beta = 0.4$
\cite{Stauffer}. In fact, it is believed that $\beta$ depends only
on the dimension of the space and not on the particular lattice used to 
tile the space, a phenomenon called universality. Plugging in our
values for $p$, $p_c$, we get that $0.24$ fraction of lattice points
belong to small clusters. This means that $0.26$ fraction of lattice
points belong to the unique giant cluster. The remaining $1 - p = 0.5$
fraction of sites is unoccupied as is to be expected.

Percolation theory also tells us that the number of finite clusters
$N_f$ is governed by another critical exponent $\alpha$ 
\cite{Stauffer}:
\begin{equation}
\label{eq:NFinite1}
N_f \approx 0.01 N_L (p - p_c)^{2-\alpha}.
\end{equation}
For the simple cubic lattice in three dimensions, $\alpha = -0.5$
\cite{Stauffer}. In fact, it is believed that $\alpha$ depends only
on the dimension of the space and not on the particular lattice used to 
tile the space, thanks to universality. Plugging in our
values for $p$, $p_c$, $N_L$, we get
\begin{equation}
\label{eq:NFinite2}
N_f \approx 10^{-4} N_L =
10^{-4} \left(
4.2 \times 10^{3.5}
\frac{M_\Pl (\frac{g_*(T_{\fo,\lo})}{10})^{1/4} T_{\fo,\lo}}
     {A \sigma_\lo^{1/2}}
\right)^3.
\end{equation}

The discussion in the above paragraphs leads us to conclude that the
arrangement of low subdomains inside one high domain follows the
so-called Swiss cheese phenomenon of percolation theory. There is a
unique giant positive-low subdomain plus a unique giant negative-low
subdomain inside a high domain. Interspersed within the giant positive-low
subdomain are small negative-low subdomains like holes in Swiss cheese.
Similarly, interspersed within the giant negative-low
subdomain are small positive-low subdomains. In addition, starting
from the outer surfaces of the two giant low subdomains and extending
all the way to the inner surface of the enclosing high domain is a
network of small positive and small negative low type subdomains. The
number of positive-low subdomains $N'_f$ near the inner surface of the 
enclosing high domain is thus given by
\begin{equation}
\label{eq:NPrimeFinite}
N'_f \approx N_f^{2/3} =
10^{-8/3} \left(
4.2 \times 10^{3.5}
\frac{M_\Pl (\frac{g_*(T_{\fo,\lo})}{10})^{1/4} T_{\fo,\lo}}
     {A \sigma_\lo^{1/2}}
\right)^2.
\end{equation}
For $T_{\fo,\lo} = 10^3\,\GeV$, 
$\frac{g_*(T_{\fo,\lo})}{10} = 10.675$ \cite{Husdal}. Plugging in
the values of $M_\Pl$, $\sigma_\lo$ gives us
$N'_f \approx 10^{44}$.

The pressure difference on a high type wall occurs due to the chance
excess of the surface small positive low subdomains over the surface small
negative low subdomains. This excess number $N''_f$ is given by
random fluctuation theory to be 
\begin{equation}
\label{eq:NDoublePrimeFinite}
N''_f \approx \sqrt{N'_f} =
10^{-4/3} \left(
4.2 \times 10^{3.5}
\frac{M_\Pl (\frac{g_*(T_{\fo,\lo})}{10})^{1/4} T_{\fo,\lo}}
     {A \sigma_\lo^{1/2}}
\right) =
\frac{616.5 \cdot M_\Pl (\frac{g_*(T_{\fo,\lo})}{10})^{1/4} T_{\fo,\lo}}
     {A \sigma_\lo^{1/2}}.
\end{equation}

From the above discussions,
whether the positive-low subdomains or the negative-low
subdomains are in excess inside a high type domain, is an event that 
occurs with probability very close to $0.5$.
Thus with probability very close to $0.25$ it happens that, say, 
a non-trivial majority of the positive-high domains contain a non-trivial 
excess of surface small positive-low subdomains and a non-trivial
majority of the negative-high domains contain a non-trivial excess of
surface small negative-low subdomains. 
We now study the cosmological consequences of this pure one out of four
chance event.

As the temperature $T$ cools below $T_{\fo,\lo}$, the energy
bias between positive-low
and negative-low types of subdomains due to the QCD instanton anomaly
as given by Equation~\ref{eq:eps} increases. We assume that the high
type walls continue to evolve according to the scaling limit. The
low type walls are constrained by frictional forces because they 
continue to interact with many 2HDM and SM particles, including 
somewhat heavy ones like the top quark, SM Higgs, $W$ and $Z$ bosons etc.
At temperatures below 4 GeV, the friction between the low type 
domain walls and the particles in the low type domains decreases 
significantly since
the wall is made from 2HDM Higgs fields which can interact with only
the light first and second generation quarks and leptons plus the tau
lepton that are left now. As we shall see below, the high and low walls
will be annihilated at lower temperatures of around 2 GeV. Following
Blasi et al. \cite{BlasiFrictionDomainWall}, we
assume that the radius of curvature of the low walls asymptotically
evolves as a power law 
\begin{equation}
\label{eq:Rlo}
R_{\lo}(t) = \frac{t^\lambda}{A}, 
\end{equation}
for some $0 < \lambda < 1$, for time $t > t_{\fo,\lo}$. 
The paper \cite{BlasiFrictionDomainWall} shows via simulations that the
asymptotic evolution of walls under mild friction is rather close to the
scaling limit. Hence we take the exponent $\lambda$ in our power law
to range from $0.9$--$0.99$ 
in this work. The consequence of the power law assumption is that the 
the radius of curvature of the low walls increases slower than scaling 
limit while their number inside one high domain remains the same.
This happens because the shapes of the low subdomains, and especially the
smaller low subdomains, are highly irregular like fractals
\cite{Stauffer}.

The volume pressure difference acting on a high type wall at temperature
$T$, $T < T_{\fo,\lo}$ thus becomes
\begin{equation}
\label{eq:deltaVhi}
\begin{array}{rcl}
\delta V_\hi(T) 
& = &
\delta V_\lo(T) N''_f \\
& = &
8 \times 10^{-4} \, 
\Lambda_{\QCD} m_u m_d m_s
\left(\frac{\Lambda_\QCD}{\pi T}\right)^8
\left(9 \ln\left(\frac{\pi T}{\Lambda_\QCD}\right)\right)^6 
\cdot
\frac{616.5 \cdot M_\Pl (\frac{g_*(T_{\fo,\lo})}{10})^{1/4} T_{\fo,\lo}}
     {A \sigma_\lo^{1/2}} \\
& = &
0.493 \,
\Lambda_{\QCD} m_u m_d m_s
\left(\frac{\Lambda_\QCD}{\pi T}\right)^8
\left(9 \ln\left(\frac{\pi T}{\Lambda_\QCD}\right)\right)^6 
\cdot
\frac{M_\Pl (\frac{g_*(T_{\fo,\lo})}{10})^{1/4} T_{\fo,\lo}}
     {A \sigma_\lo^{1/2}}.
\end{array}
\end{equation}

At a later time $t_{\ann,\hi} > t_{\fo,\lo}$ and correspondingly lower 
temperature $T_{\ann,\hi} < T_{\fo,\lo}$, the high walls experience
instability when the volume pressure difference between high domains 
starts becoming comparable to the surface tension inside the high
walls. The equation for this is given by \cite{Saikawa}:
\[
\delta V_\hi(T_{\ann,\hi}) =
C_\ann \frac{\sigma_\hi}{R_{\hi}(t_{\ann,\hi})} =
C_\ann \frac{\sigma_\hi A}{t_{\ann,\hi}},
\]
where $C_\ann$ is a constant taking values between 2 and 5. We shall
take $C_\ann = 2$ in this work.
This implies that \cite{Saikawa}
\begin{equation}
\label{eq:tannhi}
t_{\ann,\hi} =
C_\ann \frac{\sigma_\hi A}{\delta V_\hi(t_{\ann,\hi})} =
(6.58 \times 10^{-25}\,s)\,
C_\ann A (\frac{\sigma_\hi}{\mathrm{GeV}^3})
(\frac{\delta V_\hi(t_{\ann,\hi})}{\mathrm{GeV}^4})^{-1}.
\end{equation}
Since all this is occuring in the radiation dominated era,
plugging Equations~(\ref{eq:deltaVhi}), (\ref{eq:tvsT}) into
Equation~(\ref{eq:tannhi}) gives us
\begin{eqnarray*}
T_{\ann,\hi} 
& = &
3.41 \times 10^{8.5}~\mbox{GeV} \; C_{\mathrm{ann}}^{-1/2} A^{-1/2}
\left(\frac{g_*(T_{\ann,\hi})}{10}\right)^{-1/4}
(\frac{\sigma_\hi}{\mbox{GeV}^3})^{-1/2}
(\frac{\delta V_\hi(t_{\ann,\hi})}{\mbox{GeV}^4})^{1/2} \\
& = &
(7.57 \times 10^{8}~\mbox{GeV}) \; 
\frac{C_{\mathrm{ann}}^{-1/2} A^{-1} T_{\fo,\lo}^{1/2} M_\Pl^{1/2}}
     {\sigma_\lo^{1/4}} 
\left(\frac{g_*(T_{\ann,\hi})}{10}\right)^{-1/4}
\left(\frac{g_*(T_{\fo,\lo})}{10}\right)^{1/8} 
(\frac{\sigma_\hi}{\mbox{GeV}^3})^{-1/2} \\
&   &
~~~~~~~~~~
(\Lambda_{\QCD} m_u m_d m_s)^{1/2}
\left(\frac{\Lambda_\QCD}{\pi T_{\ann,\hi}}\right)^4
\left(9 \ln\left(\frac{\pi T_{\ann,\hi}}{\Lambda_\QCD}\right)\right)^3.
\end{eqnarray*}

We take
the parity breaking scale $M_P$ to range from a lower bound of
$10^6$ GeV, as discussed in Section~\ref{sec:PLRSM}, to an upper bound of
$10^7$ GeV. As mentioned earlier,
$\sigma_\hi = M_P^3$, $\sigma_\lo = 10^6\,\mathrm{GeV}^3$.  
As we shall see shortly, both high and low type walls will annihilate in
the temperature range $0.5$--$5$ GeV. In this range,
$\frac{g_*(T_{\ann,\hi})}{10} \approx 6$.
Plugging into the above equation the values 
$C_\ann = 2$, $A = 0.8$, $\Lambda_\QCD = 0.34\,\mathrm{GeV}$,
$M_\Pl = 2.435 \times 10^{18}\,\mathrm{GeV}$, 
$\frac{g_*(T_{\fo,\lo})}{10} = 10.675$ since 
$T_{\fo,\lo} = 1000\,\mathrm{GeV}$, we get
\begin{eqnarray*}
\frac{T_{\ann,\hi}^5}
     {\left(
      \ln\left(\frac{\pi T_{\ann,\hi}}{\Lambda_\QCD}\right)
      \right)^3 
     }
& = &
9^3 \cdot
(7.57 \times 10^{8}) 
\frac{C_{\mathrm{ann}}^{-1/2} A^{-1} T_{\fo,\lo}^{1/2} M_\Pl^{1/2}}
     {\sigma_\lo^{1/4}} 
\left(\frac{g_*(T_{\ann,\hi})}{10}\right)^{-1/4}
\left(\frac{g_*(T_{\fo,\lo})}{10}\right)^{1/8} 
(\frac{\sigma_\hi}{\mbox{GeV}^3})^{-1/2} \\
&   &
~~~~~~~~~~
\frac{\Lambda_{\QCD}^{4.5} (m_u m_d m_s)^{1/2}}{\pi^4} \\
& = &
9^3 \cdot
(7.57 \times 10^{8}) \cdot 10.675^{1/8} \cdot 6^{-1/4} \\
&   &
~~~~~~~~~~
\frac{0.34^{4.5} \cdot ((2.16) (4.67) (93.4))^{1/2} 10^{-9/2} \cdot 
      (10^3)^{1/2} \cdot (2.435 \times 10^{18})^{1/2}
     }
     {2^{1/2} \cdot (0.8)^{1/2} \cdot \pi^4 \cdot (10^6)^{1/4} M_P^{3/2}}\\
& = &
\frac{4.54 \times 10^{13}}{M_P^{3/2}}.
\end{eqnarray*}
For $M_P = 10^6\,\GeV$, we get
$
\frac{T_{\ann,\hi}^5}
     {\left(
      \ln\left(\frac{\pi T_{\ann,\hi}}{0.34}\right)
      \right)^3 
     } =
4.54 \times 10^4.
$
For $M_P = 10^7\,\GeV$, we get
$
\frac{T_{\ann,\hi}^5}
     {\left(
      \ln\left(\frac{\pi T_{\ann,\hi}}{0.34}\right)
      \right)^3 
     } =
1.44 \times 10^3.
$
Solving these transcendental equations numerically gives the
temperature $T_{\ann,\hi}$ of the instability of the high walls to 
range from 23.4 GeV to 10.7 GeV. 

The low type subdomain walls experience instability at a
time $t_{\ann,\lo} > t_{\fo,\lo}$ corresponding to temperature
$T_{\ann,\lo} < T_{\fo,\lo}$ when the 
volume pressure difference between low domains starts becoming comparable 
to the surface tension inside the low walls.
The equation for this is given by \cite{Saikawa}:
\[
\delta V_\lo(T_{\ann,\lo}) =
C_\ann \frac{\sigma_\lo}{R_{\lo}(t_{\ann,\lo})} =
C_\ann \frac{\sigma_\lo A}{t_{\ann,\lo}^\lambda}.
\]
This implies that \cite{Saikawa}
\begin{equation}
\label{eq:tannlo}
t_{\ann,\lo} =
\left(
\frac{\sigma_\lo C_\ann A}{\delta V_\lo(t_{\ann,\lo})}
\right)^{1/\lambda} =
(6.58 \times 10^{-25}\,s)\,
(C_\ann A)^{1/\lambda} 
(\frac{\sigma_\lo}{\mathrm{GeV}^3})^{1/\lambda}
(\frac{\delta V_\lo(t_{\ann,\lo})}{\mathrm{GeV}^4})^{-1/\lambda}.
\end{equation}
The pressure difference between the low type domains at temperature 
$T_{\ann,\lo}$ is given by
\begin{equation}
\label{eq:deltaVlo}
\delta V_\lo(T_{\ann,\lo}) =
8 \times 10^{-4} \, 
\Lambda_{\QCD} m_u m_d m_s
\left(\frac{\Lambda_\QCD}{\pi T_{\ann,\lo}}\right)^8
\left(9 \ln\left(\frac{\pi T_{\ann,\lo}}{\Lambda_\QCD}\right)\right)^6.
\end{equation}
Since all this is occuring in the radiation dominated era,
plugging Equations~(\ref{eq:deltaVlo}), (\ref{eq:tvsT}) into
Equation~(\ref{eq:tannlo}) gives us
\begin{eqnarray*}
T_{\ann,\lo} 
& = &
3.41 \times 10^{8.5}~\mbox{GeV} \; 
C_{\ann}^{-1/(2\lambda)} A^{-1/(2\lambda)}
\left(\frac{g_*(T_{\ann,\lo})}{10}\right)^{-1/4}
(\frac{\sigma_\lo}{\mbox{GeV}^3})^{-1/(2\lambda)}
(\frac{\delta V_\lo(t_{\ann,\lo})}{\mbox{GeV}^4})^{1/(2\lambda)} \\
& = &
(3.41) (8^{1/(2\lambda)}) (10^{8.5 - \frac{2}{\lambda}}) \;
(C_{\ann}^{-1/(2\lambda)} A^{-1/(2\lambda)})
\left(\frac{g_*(T_{\ann,\lo})}{10}\right)^{-1/4}
\sigma_\lo^{-1/(2\lambda)} \\
&   &
~~~~~~~~~~
(\Lambda_{\QCD} m_u m_d m_s)^{1/(2\lambda)}
\left(\frac{\Lambda_\QCD}{\pi T_{\ann,\hi}}\right)^{4/\lambda}
\left(
9 \ln\left(\frac{\pi T_{\ann,\hi}}{\Lambda_\QCD}\right)
\right)^{3/\lambda}.
\end{eqnarray*}
Plugging in the same values as before, we get
\begin{eqnarray*}
\frac{T_{\ann,\lo}^{1 + \frac{4}{\lambda}}}
     {\left(
      \ln\left(\frac{\pi T_{\ann,\lo}}{\Lambda_\QCD}\right)
      \right)^{3/\lambda}
     }
& = &
9^{3/\lambda} (3.41) 6^{-1/4} \cdot 8^{1/(2\lambda)}
(10^{8.5 - \frac{2}{\lambda} - \frac{9}{2\lambda} - \frac{3}{\lambda}})
\cdot
\frac{0.34^{9/(2\lambda)} ((2.16) (4.67) (93.4))^{1/(2\lambda)}}
     {2^{1/(2\lambda)} \cdot (0.8)^{1/(2\lambda)} \cdot \pi^{4/\lambda}} \\
& = &
(2.18 \times 10^{8.5})
\left(
\frac{9^3 \cdot 8^{1/2} \cdot 10^{-9.5} \cdot (0.34)^{4.5} 
      ((2.16) (4.67) (93.4))^{1/2}
     }
     {2^{1/2} \cdot (0.8)^{1/2} \cdot \pi^4}
\right)^{1/\lambda} \\
& = &
(2.18 \times 10^{8.5})
(4 \times 10^{-9.5})^{1/\lambda}.
\end{eqnarray*}
At the higher friction end of the evolution of the low walls when
$\lambda = 0.9$, we get
$
\frac{T_{\ann,\lo}^{5.44}}
     {\left(
      \ln\left(\frac{\pi T_{\ann,\lo}}{0.34}\right)
      \right)^{3.33}
     } =
0.09.
$
At the lower friction end of the evolution of the low walls when
$\lambda = 0.99$, we get
$
\frac{T_{\ann,\lo}^{5.04}}
     {\left(
      \ln\left(\frac{\pi T_{\ann,\lo}}{0.34}\right)
      \right)^{3.03}
     } =
0.79.
$
Solving these transcendental equations numerically gives the
temperature $T_{\ann,\lo}$ of the instability of the low walls to 
range from 1.09 GeV to 1.77 GeV. 

The above analysis shows that for a range of values $10^6$--$10^7$ GeV 
of the parity breaking scale $M_P$, and for a range of values 
$0.9$--$0.995$
of the power in the friction dominated evolution of the low walls,
both high and low walls experience instability within  one order of 
magnitude of temperature range $1$--$25$ GeV. So both types of walls
annihilate at almost the same time resulting finally in a single domain
which is our familiar left symmetric SM current universe.

We have thus shown that both high type and low type domain walls can
disappear at almost the same time well before the advent of BBN, which 
occurs at a temperature of
$1~\mathrm{MeV}$ \cite{BBNPDG}. Hence, these walls do not leave any
discernible signature in the cosmic microwave background radiation (CMBR)
and do not contradict the Zel'dovich-Kobzarev-Okun bound \cite{ZKOBound} 
of maximum
domain wall tension being around few MeV based on observations of the
CMBR.

\section{Gravitational waves from collapse of domain walls}
\label{sec:GWsignals}
The almost simultaneous collapse of the two types of domain walls at
temperatures within one order of magnitude around $T_\ann = 2\,\GeV$ 
results in 
primordial gravitational waves. Detection of these 
primordial waves, distinct from the presently detected astrophysical 
gravitational waves, would be a breakthrough discovery giving significant 
clues about
the origin and evolution of the very early universe before BBN
\cite{Caldwell:2022qsj}. Several earlier works 
\cite{Hiramatsu2010GravitationalWalls, Kadota2015GravitationalModel}
have specifically studied the primordial
gravitational waves produced by collapsing domain walls
in this regard.

The peak frequency of gravitational waves caused by the collapse of
domain walls red shifted to the present time $t_0$ is given 
by \cite{Saikawa}
\begin{equation}
\label{eq:peakfreq}
f_{\mathrm{peak}} =
1.1 \times 10^{-7}~\mbox{Hz}~
\left(\frac{g_*(T_\ann)}{10}\right)^{1/2}
\left(\frac{g_{*s}(T_\ann)}{10}\right)^{-1/3}
\left(\frac{T_\ann}{\mbox{GeV}}\right),
\end{equation}
where $g_{*s}(T_\ann)$ is the number of relativistic degrees 
of freedom for the entropy density at temperature $T_\ann$.
The peak energy density spectrum of the waves from the collapse of
the high walls at the present time 
is given by \cite{Saikawa}
\begin{equation}
\label{eq:peakenergyhi}
\Omega_{\mathrm{gw}} h^2(t_0)_{\mathrm{peak},\hi} =
7.2 \times 10^{-44} \tilde{\epsilon}_{\mathrm{gw}} A^2 
\left(\frac{g_{*s}(T_\ann)}{10}\right)^{-4/3}
\left(\frac{\sigma_\hi}{\mbox{GeV}^3}\right)^{2}
\left(\frac{T_\ann}{\mbox{GeV}}\right)^{-4},
\end{equation}
where $\tilde{\epsilon}_{\mathrm{gw}} \approx 0.7 \pm 0.4$ 
and $A \approx 0.8 \pm 0.1$ for 
$Z_2$-domain walls annhilated in the radiation era.
Similarly, the peak energy density spectrum of the waves from the 
collapse of the low walls at the present time 
is given by 
\begin{equation}
\label{eq:peakenergylo}
\Omega_{\mathrm{gw}} h^2(t_0)_{\mathrm{peak},\lo} =
7.2 \times 10^{-44} \tilde{\epsilon}_{\mathrm{gw}} A^2 
\left(\frac{g_{*s}(T_\ann)}{10}\right)^{-4/3}
\left(\frac{\sigma_\lo}{\mbox{GeV}^3}\right)^{2}
\left(\frac{T_\ann}{\mbox{GeV}}\right)^{-4},
\end{equation}

At annihilation temperatures $T_\ann$ 
ranging from $1$--$10$ GeV,
both $\frac{g_*(T_\ann)}{10}$ and $\frac{g_{*s}(T_\ann)}{10}$ are around
6 \cite{Husdal}. 
The peak frequency of gravitational waves from both types of collapsing
walls for annihilation temperatures of $1$--$10$ GeV is around
\begin{equation}
\label{eq:peakfreq}
f_{\mathrm{peak}}(10\,\GeV) =
1.48 \times 10^{-6}~\mbox{Hz},
~~~
f_{\mathrm{peak}}(1\,\GeV) =
1.48 \times 10^{-7}~\mbox{Hz}.
\end{equation}
The peak energy density of the gravitational waves from the collapse of
the high walls for parity breaking scales of $10^6$--$10^7$ GeV and
annihilation temperatures of $1$--$10$ GeV
is around
\begin{equation}
\label{eq:peakamplitude}
\begin{array}{c}
\Omega_{\mathrm{gw}} h^2(t_0)_{\mathrm{peak},\hi}(10^6, 1) = 
2.96 \times 10^{-9},
~~~
\Omega_{\mathrm{gw}} h^2(t_0)_{\mathrm{peak},\hi}(10^6, 10) = 
2.96 \times 10^{-13}, \\
\Omega_{\mathrm{gw}} h^2(t_0)_{\mathrm{peak},\hi}(10^7, 1) = 
2.96 \times 10^{-3},
~~~
\Omega_{\mathrm{gw}} h^2(t_0)_{\mathrm{peak},\hi}(10^7, 10) = 
2.96 \times 10^{-7}.
\end{array}
\end{equation}
The peak energy density of the gravitational waves from the collapse of
the low walls is insignificant in comparison to the high walls 
because $\sigma_\lo \ll \sigma_\hi$.

These frequencies are about 7 to 8 orders of
magnitude below what LIGO-VIRGO \cite{LIGO} can detect today, 
and what future
ground based gravitational wave telescopes like Einstein Telescope
\cite{Punturo:2010zz}
are designed to detect. They are also about 6 orders of magnitude
below what future space based gravitational wave detectors like
DECIGO \cite{Yagi:2011wg} are designed to detect.
However, proposed
pulsar timing array based gravitational wave detectors like 
SKA \cite{SKA} as well as existing pulsar timing array based detectors like
NANOGrav \cite{NANOGRAV:2023} are expressely designed to detect 
frequencies in the $10^{-9}$--$10^{-7}$ Hz range.
Very recently, NANOGrav has published strong evidence for a stochastic 
gravitational wave background with frequency around $10^{-7}$ Hz 
($1\,\mathrm{yr}^{-1}$) and
integrated energy density of $9.3 \times 10^{-9}$ based on their 
15 year data set \cite{NANOGRAV:2023}. 
Their results are consistent with astrophysical expectations for a signal 
from a population of supermassive black hole binaries, although more 
exotic cosmological and astrophysical sources like domain wall collapse
cannot be excluded.
The NANOGrav results immediately rule out
our DW collapse model with a higher parity breaking scale of $10^7$ GeV. 
However, our DW collapse model with a parity breaking scale of
$10^6$ GeV remains entirely consistent with the NANOGrav results. 
Further data collection and improvements in pulsar timing array based
detectors in the near future can thus serve as a strong experimental
test of our model.

\section{Conclusion and discussion}
\label{sec:conclusion}
In this work we have addressed the problem of domain wall creation and 
removal for a class 
of BSM theories with the common feature of discrete
$Z_2$-symmetry breaking. 
We have shown that the seminal PLRSM model of Chang, Mohapatra and Parida
\cite{Parida}, which was proposed to show that it is possible to 
break discrete
parity without breaking left right gauge symmetry, is also one such
$Z_2$-symmetry breaking theory. In particular, PLRSM has a hitherto
unnoticed consequence of creating a domain wall at the time of discrete
parity breaking, before gauge symmetry breaking. 

Without the proportionality condition on the Yukawa couplings, PLRSM
and LRSM have a domain wall problem. The walls are of the high type
as discussed in the previous sections of this paper.
One needs to add $Z_2$-symmetry breaking non-renormalisable operators
to the Lagrangian as in \cite{RS:1994}
to remove these walls successfully before they contradict with 
standard cosmology.

Our work shows that however, when the proportionality constraint on
the Yukawa couplings holds, PLRSM and LRSM reduce
to Type II 2HDM at lower energies. We then show that there 
is indeed a mechanism via the formation of low type plus
high type domain walls combined with an energy bias due to QCD
instanton anomaly and random fluctuations by which all domain 
walls disappear well before BBN.

Our domain wall removal mechanism starts 
by showing that a $Z_2$-domain formed at a high energy scale of discrete
parity breaking decomposes further into many subdomains 
due to another $Z_2$-symmetry breaking
present in the Type II 2HDM model with spontaneous CP violation at a 
much lower scale.  A QCD instanton vertex anomaly first 
identified by \cite{PreskillTrivedi}, on combining with the percolation
properties of the distribution
of subdomains within a domain statistical and the
properties of random fluctuations, creates
enough pressure difference to remove the both high and low types of walls
almost simultaneously well before 
BBN.  The probability of getting a favourable excess of low type 
subdomains near the surface of a high type domain due to
random fluctations, which is key  for the removal of the
walls, is very close to $0.25$.

We have performed detailed calculations of our main idea for the PLRSM
model with parity breaking scale ranging from $10^6$--$10^7$ GeV and
friction power law exponent ranging from $0.9$--$0.99$.
The peak frequency of gravitational waves resulting from both types of 
wall collapse ranges from $10^{-6}$ to $10^{-7}~\mbox{Hz}$. 
This frequency band is sensitive  
to pulsar timing array based experiments such as SKA and NANOGrav. The
recent NANOGrav results rule out our DW collapse model for higher
values of parity breaking scale above $10^7$ GeV. Our DW collapse model
with parity breaking scales below $10^7$ GeV  remains consistent with 
the current NANOGrav results and has a good
chance of being seriously tested in future pulsar timing based experiments.
Our work gives further impetus for gravitational wave 
astronomy in the $10^{-7}$--$10^{-6}$ Hz band in the near future. 

Complementary to the possibility of primordial gravitational waves arising
from domain wall collapse is the possibility of primordial gravitational 
waves arising from collapse of cosmic strings in the very early universe
and the scope of their detecting by existing and planned experiments,
see e.g.  \cite{AuclairGW, CuiGW, EllisGW, BlasiGW}. Many interesting
BSM models predict such strings and gravitational waves may offer a
novel way of studying them experimentally, see 
e.g. \cite{DrorLepto, BlasiLepto}.

Proceeding in a very different direction,
we observe via the results of Chen et al. \cite{Chen} that upcoming 
experiments for measuring 
the electric dipole moment (EDM) of the electron will constrain some
spontaneous CP violating parameters of the underlying Type II 2HDM
model involved in the subdomain formation. In particular, the
CP violating mixing angle $\alpha_c$ of Chen et al. will be similarly
constrained to lie below $10^{-2}$ for reasonable values of the
mass splitting between the two neutral heavy Higgs bosons of the Type II
2HDM due to the upper bound on electron EDM obtained by the ACME2
experiment \cite{ACME2}. A negative result in the upcoming ACME3 
experiment will further constrain $\alpha_c$ to lie below $10^{-4}$.
However, unlike the findings of Chen et
al., our domain wall removal mechanism does not need any explicit
CP violating term in the Type II 2HDM Higgs potential; instead it exploits
of the QCD anomaly of Preskill et al. The main insight provided by our
work is the increased importance and ubiquitousness of the 2HDM,
as well as the demonstration that {\em successful domain wall removal is
possible with only spontaneous CP violation in the 2HDM}.

Though our detailed calculations have been carried out for the PLRSM 
model only, we believe
that the main idea of domain wall removal via the formation of
Type II 2HDM subdomains at a lower energy scale, followed by the
creation of a certain favourable excess of subdomains statistically
with probability of around $0.25$, and finally followed by the
collapse of the subdomain walls due to the QCD anomaly is sufficiently 
robust. Our work leads one to the tantalising idea
that {\em the universe we observe is a purely chance  survival of one out 
of four distinct possibilities}. Of course if there were soft
explicit $Z_2$-symmetry breaking terms in the Lagrangians involved, the
domain wall collapse becomes even easier.
Our main result is agnostic with respect this provenance.

\begin{acknowledgments}
PB was supported by a Women 
Scientists-A Fellowship from the Department of Science and Technology.
UAY is supported
by an IIT Bombay Institute Chair Professorship. We thank the anonymous
referees of a previous draft for their extremely helpful 
suggestions.
\end{acknowledgments}

\bibliography{Parida}

\end{document}